\documentclass[aps,prl,reprint,showpacs,byrevtex]{revtex4-2}
\let\oldhat\hat
\renewcommand{\hat}[1]{\oldhat{\mathbf{#1}}}
\usepackage{titlesec}
\usepackage{array}

\usepackage{times,mathptm}
\usepackage[dvips]{graphicx,color}

\begin{document}
\title{Driving Mechanism and Dynamic Fluctuations of Charge Density Waves in the Kagome Metal ScV$_6$Sn$_6$}
\author{Shuyuan Liu$^{1,2}$, Chongze Wang$^2$, Shichang Yao$^2$, Yu Jia$^{1,3}$, Zhenyu Zhang$^{4*}$, and Jun-Hyung Cho$^{2*}$}
\affiliation{$^1$Joint Center for Theoretical Physics, School of Physics and Electronics, Henan University, Kaifeng 475004, People's Republic of China \\
$^2$Department of Physics and Research Institute for Natural Science, Hanyang University, 222 Wangsimni-ro, Seongdong-Ku, Seoul 04763, Republic of Korea \\
$^3$Key Laboratory for Special Functional Materials of the Ministry of Education, Henan University, Kaifeng 475004, People's Republic of China \\
$^4$International Center for Quantum Design of Functional Materials (ICQD), Hefei National Laboratory for Physical Sciences at Microscale, and Synergetic Innovation Center of Quantum Information and Quantum Physics, University of Science and Technology of China, Hefei 230026, China}
\date{\today}

\begin{abstract}
In contrast to the AV$_3$Sb$_5$ (A = K, Rb, Cs) family exhibiting the ubiquitous 2${\times}$2${\times}$2 CDW order, only ScV$_6$Sn$_6$ in the RV$_6$Sn$_6$ (R = Sc, Y, La) family displays the unusual CDW with a $\sqrt{3}{\times}\sqrt{3}$ in-plane ordering and a tripling of the unit cell along the $c$-axis. Here, using first-principles density-functional theory calculations, we show that both the $\sqrt{3}{\times}\sqrt{3}{\times}$2 and $\sqrt{3}{\times}\sqrt{3}{\times}$3 CDW orderings can be driven by a Jahn-Teller-like effect where the Sn atoms residing in the kagome bilayers partially undergo delicately different schemes of interlayer dimerization, accompanied by charge redistribution between such Sn atoms and band gap opening. Counterintuitively, whereas the $\sqrt{3}{\times}\sqrt{3}{\times}$2 phase is energetically more stable than $\sqrt{3}{\times}\sqrt{3}{\times}$3, the latter is thermodynamically stabilized above the CDW transition temperature $T_{\rm CDW}$ by its higher configurational entropy contributed by degenerate fluctuating phases, and is kinetically selected below $T_{\rm CDW}$, as corroborated by experimental observation of the accompanying first-order phase transition. Our findings reveal the order-disorder nature of the $\sqrt{3}{\times}\sqrt{3}{\times}$3 CDW phase transition in ScV$_6$Sn$_6$, with broader implications for understanding charge orderings and CDW fluctuations in other kagome metals.
\end{abstract}
\pacs{}
\maketitle


$Introduction$.$-$ The two-dimensional (2D) kagome lattice consisting of corner-sharing triangles has attracted tremendous attention due to its unique electronic structure involving a flat band, Dirac cone, and van Hove singularities (VHSs)~\cite{AHC-Fe3Sn2-Nat2018, KagomevHs-PRB2021, FeSn_PRB}. So far, many experimental and theoretical works in different kagome compounds have presented a wide variety of novel quantum phenomena like anomalous Hall effect~\cite{AHC-Mn3Sn-Nat2015, AHC-Mn3Ge-Sci2016, AHC-Co3Sn2S2-Nat2018}, spin liquid~\cite{SpinLiquid-Nat2010, SpinLiquid-Sci2011}, flatband ferromagnetism~\cite{Fe3Sn2-USTC-Zeng}, charge density wave (CDW)~\cite{AV3Sb5-PRM2019, KV3Sb5-chrialCDW-Nat.Mat2021}, and superconductivity~\cite{CsV3Sb5-Z2-PRL2020, KV3Sb5_SC_Z2_PRM2021, RbV3Sb5-SC-CPL2021}. Specifically, a family of V-based kagome metals AV$_3$Sb$_5$ (A = K, Rb, Cs) possesses nontrivial band topology and correlated many-body states such as the unconventional CDW order breaking both time-reversal and rotational symmetries~\cite{KV3Sb5-AHC-AS2020, CsV3Sb5-AHC_SC-PRB2021}, electronic nematicity~\cite{en1,en2}, and superconductivity~\cite{CsV3Sb5-CDW_SC-NC2021, CsV3Sb5-SC_CDW-PRL2021, CsV3Sb5_SC_PRB2021, chongze-PRM}. Considering that the AV$_3$Sb$_5$ compounds have the saddle points of linearly dispersive Dirac bands at three inequivalent $M$ points near the Fermi level $E_F$, the nesting of such VHSs was proposed to be a driving mechanism for the CDW formation via electron-electron interaction~\cite{Peiers, Kohn-PRL1959, KagomeFS-PRL2013, KagommeCDW-PRB2013}. However, the origin of CDW in AV$_3$Sb$_5$ is still under debate, together with momentum-dependent electron-phonon coupling or phonon softening~\cite{KV3Sb5-CDW_Gap, RbV3Sb5-CDW_Gap, CsV3Sb5-CDW_Gap1, CsV3Sb5-CDW_Gap2,CsV3Sb5_natcomm}. To resolve this debate, more extensive studies of the CDW formation in other V-based kagome metals are highly desired.

Recently, among the new nonmagnetic RV$_6$Sn$_6$ (R = Sc, Y, La) family consisting of V kagome bilayers [see Fig. 1(a)], only ScV$_6$Sn$_6$ undergoes a first-order CDW transition at $T_{\rm CDW}$ $\approx$ 92 K~\cite{ScV6Sn6_CDW_PRL2022}. Here, ScV$_6$Sn$_6$ was experimentally observed to exhibit a $\sqrt{3}$${\times}$$\sqrt{3}$${\times}$3 CDW order, which is distinct from the 2${\times}$2${\times}$2 CDW order in AV$_3$Sb$_5$~\cite{AV3Sb5-PRM2019, AV3Sb5_chongze}. Such entirely different wave lengths and orientations of the CDW orders between ScV$_6$Sn$_6$ and AV$_3$Sb$_5$ may provide a new route to gain insights into the origin of the CDW order in V-based kagome metals. Specifically, while the CDW order in AV$_3$Sb$_5$ accompanies large in-plane displacements of the V atoms~\cite{AV3Sb5_chongze}, that in ScV$_6$Sn$_6$ is dominated by displacements of the Sc and Sn atoms along the $c$-axis~\cite{YBH_PRL_2023}. These unusual features of the CDW order in ScV$_6$Sn$_6$ challenge the nesting scenario because the scattering between the $M$-saddle points near $E_F$ cannot be responsible for a driving force of the $\sqrt{3}$${\times}$$\sqrt{3}$${\times}$3 periodic lattice distortion.

For ScV$_6$Sn$_6$, diffuse scattering (DS) and inelastic x-ray scattering (IXS) experiments~\cite{IXS_166} reported the presence of an order-disorder CDW phase transition at $T_{\rm CDW}$, where the disordered phase containing a short-range $\sqrt{3}$${\times}$$\sqrt{3}$${\times}$2 CDW order was suppressed and replaced by a long-range $\sqrt{3}$${\times}$$\sqrt{3}$${\times}$3 CDW order. Another DS and IXS experiments~\cite{Softening166} observed that the low energy longitudinal phonon with propagation wavevector $q^*$ = (${1 \over 3}$, ${1 \over 3}$, ${1 \over 2}$) collapses at ${\sim}$98 K due to the electron-phonon interaction. Here, the collapse of a soft mode at $q^*$ is driven by a softening of a flat phonon plane at $k_z$ = ${\pi}$, characterized by an out-of-plane vibration of the trigonal Sn atoms. However, the long-range charge order does not emerge at $q^*$ but at a different wavevector $q_s$ = (${1 \over 3}$, ${1 \over 3}$, ${1 \over 3}$). It is interesting to compare the characteristics of order-disorder CDW phase transition between ScV$_6$Sn$_6$ and AV$_3$Sb$_5$. For AV$_3$Sb$_5$, recent DS and IXS experiments~\cite{firstorder135-1,firstorder135-2,fluctuations135} also reported a first-order CDW phase transition that features an order-disorder transformation type with strong CDW fluctuations above $T_{\rm CDW}$. Note that inelastic x-ray, neutron, and Raman scattering experiments for AV$_3$Sb$_5$~\cite{firstorder135-1,firstorder135-2,fluctuations135} have observed the absence of the lattice collapse at $T_{\rm CDW}$. Therefore, both ScV$_6$Sn$_6$~\cite{IXS_166,disorder1} and AV$_3$Sb$_5$~\cite{firstorder135-1,firstorder135-2,fluctuations135} undergo the first-order, order-disorder CDW phase transition, but their characteristics are apparently distinct from each other: i.e., ScV$_6$Sn$_6$ features a first-order transition with phonon softening while AV$_3$Sb$_5$ exhibits CDW fluctuations with the absence of phonon softening. Specifically, the temperature evolution of the integrated IXS intensities in ScV$_6$Sn$_6$~\cite{IXS_166} showed that the maximum peak intensity of the $\sqrt{3}{\times}\sqrt{3}{\times}$3 CDW order below $T_{\rm CDW}$ is at least 3 orders of magnitude larger than that of the short-range $\sqrt{3}{\times}\sqrt{3}{\times}$2 CDW order above $T_{\rm CDW}$. This huge different IXS intensities of two CDW orders together with the inconsistency between the soft $q^*$-phonon and $q_s$-CDW wavevectors are puzzling. Microscopic understanding of such an unconventional CDW phase transition in ScV$_6$Sn$_6$ is yet to be identified.

\begin{figure}[h!t]
\includegraphics[width=8.5cm]{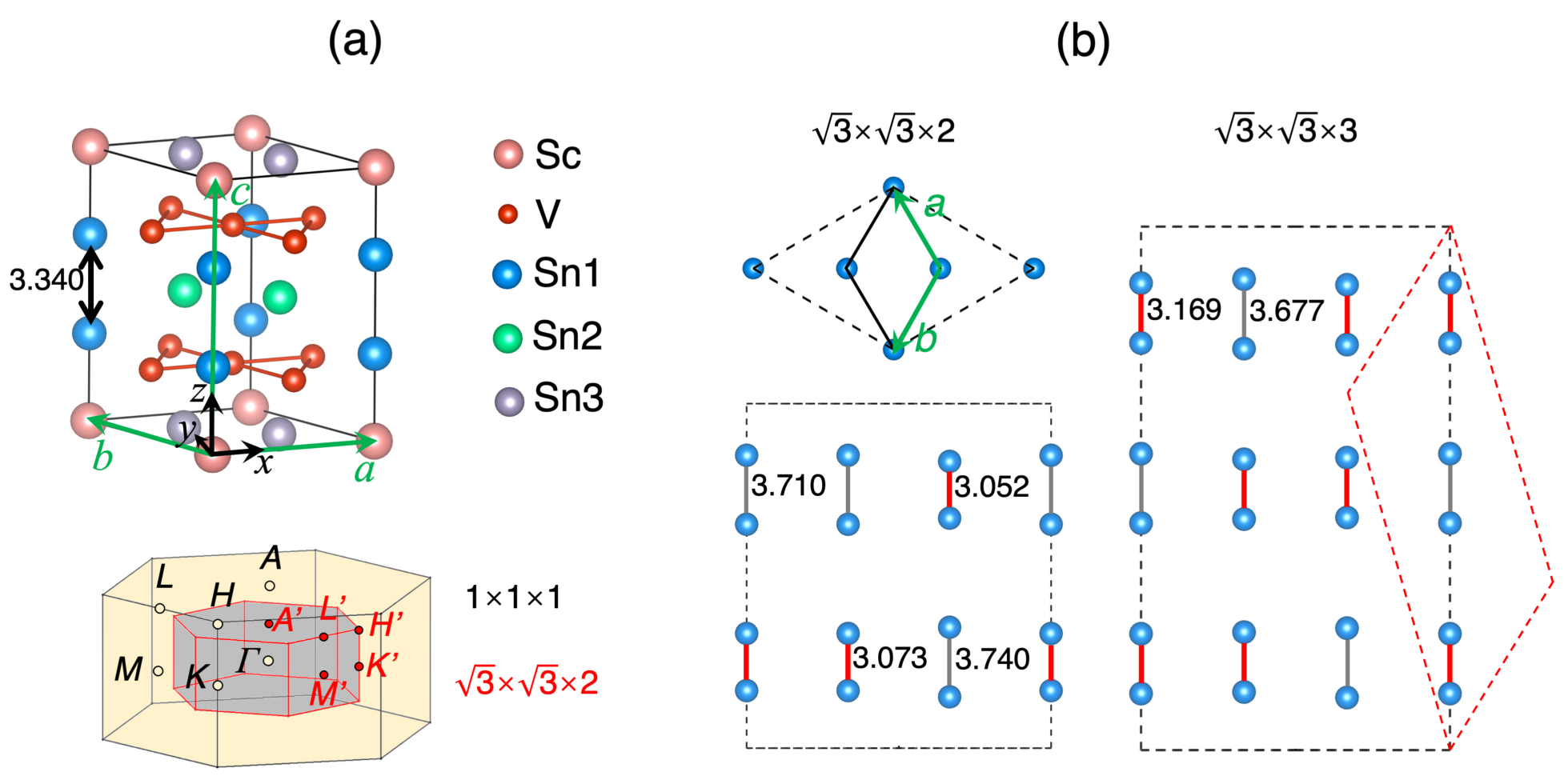}
\caption{Atomic structure of pristine and CDW phases of ScV$_6$Sn$_6$. (a) Optimized structure of the 1${\times}$1${\times}$1 pristine phase of ScV$_6$Sn$_6$, together with its Brillouin zone. The top/side views of Sn dimers in the $\sqrt{3}{\times}\sqrt{3}{\times}$2 and $\sqrt{3}{\times}\sqrt{3}{\times}$3 phases are drawn in (b). The black dashed lines in (b) represent the hexagonal unit cells, while the red dashed line in the $\sqrt{3}{\times}\sqrt{3}{\times}$3 phase represents its primitive rhombohedral unit cell. The numbers represent the Sn1-Sn1 bond lengths in angstroms.}
\label{figure:1}
\end{figure}

In this Letter, we investigate the driving mechanism of CDW and the nature of CDW phase transition in ScV$_6$Sn$_6$ using first-principles density-functional theory (DFT) calculations. We demonstrate that both the $\sqrt{3}{\times}\sqrt{3}{\times}$2 and $\sqrt{3}{\times}\sqrt{3}{\times}$3 CDW orderings can be driven by a Jahn-Teller-like effect where the Sn atoms residing in kagome bilayers are partially dimerized along the $c$-axis together with the charge redistribution between such Sn atoms and its associated band gap opening. The $\sqrt{3}{\times}\sqrt{3}{\times}$3 phase is found to be energetically less stable than $\sqrt{3}{\times}\sqrt{3}{\times}$2, but the former becomes thermodynamically stabilized above $T_{\rm CDW}$ due to its higher vibrational, electronic, and configurational entropies compared to the latter. Furthermore, we reveal that, while the $\sqrt{3}{\times}\sqrt{3}{\times}$3 CDW order shows dynamic fluctuations between its degenerate configurations above $T_{\rm CDW}$, it is kinetically trapped below $T_{\rm CDW}$, thereby leading to the first-order, order-disorder phase transition with a release of configurational entropy at $T_{\rm CDW}$. Therefore, we conclude that the majority of disordered phase above $T_{\rm CDW}$ is composed of the dynamically fluctuating $\sqrt{3}$${\times}$$\sqrt{3}$${\times}$3 CDW order, while the minority is a local ordering of the short-range $\sqrt{3}$${\times}$$\sqrt{3}$${\times}$2 CDW order due to a relatively higher energy barrier between its degenerate configurations, as discussed below.

$Results$.$-$ We begin by optimizing the atomic structure of the 1${\times}$1${\times}$1 pristine phase of ScV$_6$Sn$_6$ using the DFT scheme~\cite{method}. Figure 1(a) shows the optimized structure of the pristine phase with the symmetry of space group $P6/mmm$, which is composed of the two V$_3$Sn1 kagome layers with the triangular Sn1 sublattice centered on the V hexagons of each kagome lattice and the Sn2 honeycomb or Sn3/Sc honeycomb/triangular layer between neighboring V$_3$Sn1 kagome layers. The electronic band structure of this pristine phase is displayed in Fig. 2(a), together with its projection onto the V 3$d$ and Sn 5$p$ orbitals (see also Fig. S1 in the Supplemental Material~\cite{SM}). We find that the bands with momentum-dependent orbital characters near $E_F$ agree well with angle-resolved photoemission spectroscopy (ARPES) and optical reflection measurements~\cite{ref_note1,ARPES166,opticalPRB}: i.e., the states along the ${\Gamma}$-$M$-$K$-${\Gamma}$ line originate from all the five V $d$ orbitals, while those along the $A$-$L$-$H$-$A$ line mainly originate from the Sn1 $p_z$ and V $d_{xz,yz}$ orbitals. The latter hybridized states in the V$_3$Sn1 kagome layers play a crucial role in the CDW formation, as discussed below. Meanwhile, two saddle points at the $M$ point close to $E_F$ (indicated by red arrows) don't produce the corresponding VHSs in the partial density of states (PDOS) [see Fig. 2(a)]. Therefore, we can say that the Fermi-surface nesting of saddle-point derived VHSs between three distinct $M$ points~\cite{ref_note2}, previously proposed in AV$_3$Sb$_5$~\cite{Peiers, Kohn-PRL1959, KagomeFS-PRL2013, KagommeCDW-PRB2013}, is not responsible for the driving mechanism of the CDW order in ScV$_6$Sn$_6$~\cite{YBH_PRL_2023}. Indeed, the electronic states along the ${\Gamma}$-$M$-$K$-${\Gamma}$ line remain intact across the CDW transition, as will be shown later.

\begin{figure}[h!t]
\includegraphics[width=8.5cm]{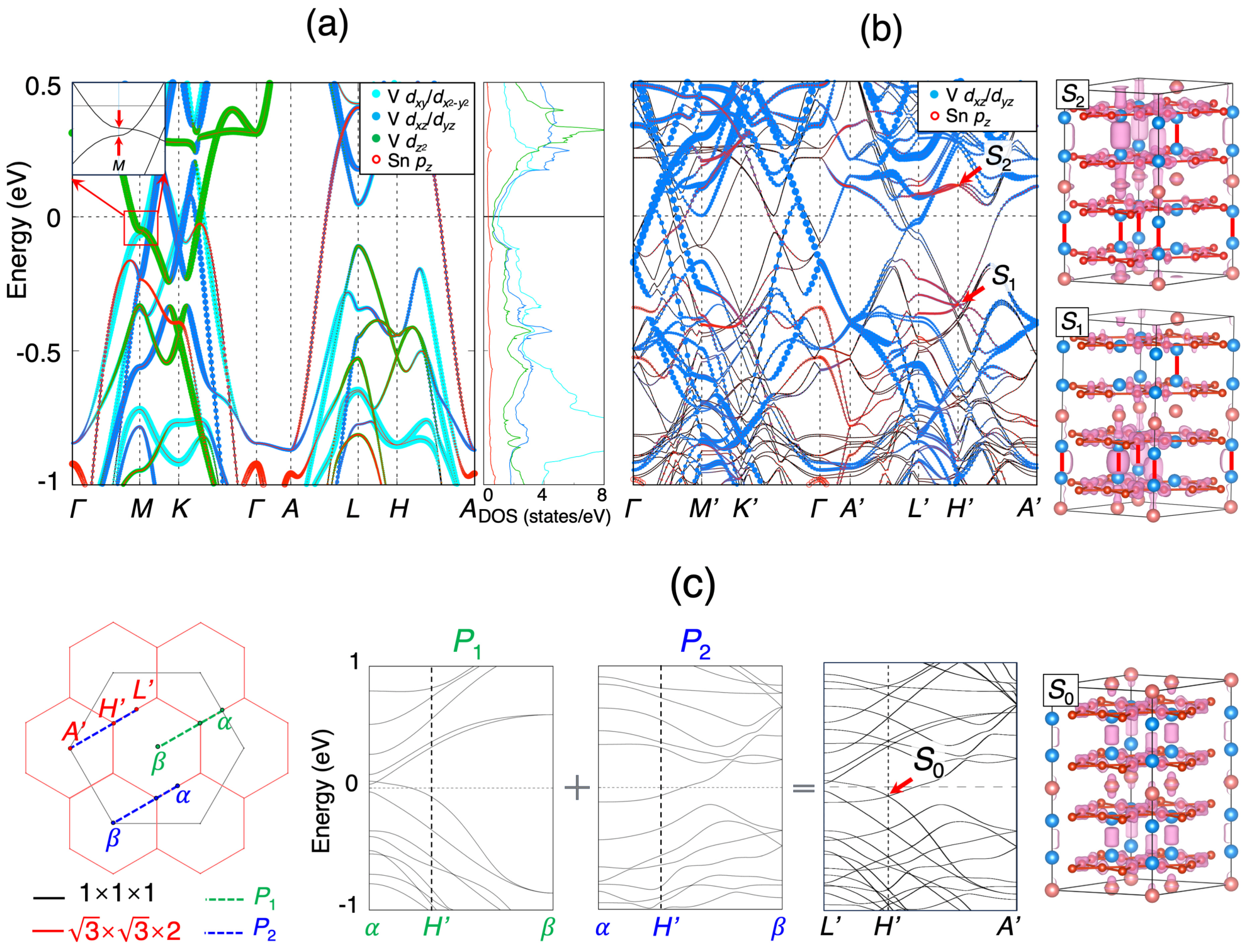}
\caption{Electronic structure of pristine and CDW phases of ScV$_6$Sn$_6$. Calculated band structures of the (a) pristine and (b) $\sqrt{3}{\times}\sqrt{3}{\times}$2 phases of ScV$_6$Sn$_6$. The bands are projected onto V 3$d$ and Sn 5$p$ orbitals, where the radii of circles are proportional to the weights of the corresponding orbitals. In (a), the PDOS and two saddle points at the $M$ point (indicated by arrows in the inset) are included. In (b), the charge characters of the $S_1$ and $S_2$ states at the $H'$ point are displayed with an isosurface of 8${\times}10^{-3}e$/{\AA}$^3$. In (c), the pristine bands along the three ${\alpha}$-${\beta}$ paths (one $P_1$ and two $P_2$) are folded to form the six-fold degenerate state $S_0$ at $H'$, the charge character of which is also displayed with an isosurface of 8${\times}10^{-3}e$/{\AA}$^3$.}
\label{figure:2}
\end{figure}

To explore the phonon instability of the pristine structure of ScV$_6$Sn$_6$, we calculate its phonon spectrum using the frozen-phonon method~\cite{phononpy1,phononpy2}. As shown in Fig. S4(a), there exist imaginary phonon modes along the entire $A$-$L$-$H$-$A$ line, the lowest one of which is located at the $H$ = (${1 \over 3}$, ${1 \over 3}$, ${1 \over 2}$) point, leading to a structural reconstruction within the $\sqrt{3}{\times}\sqrt{3}{\times}$2 unit cell. Here, the $H$-phonon mode represents the Sn1-Sn1 bond stretching vibrations along the c-axis [see Fig. S4(b)]. The resulting $\sqrt{3}{\times}\sqrt{3}{\times}$2 CDW order with the $D_{6h}$ symmetry of space group $P6/mmm$ becomes more stable than the pristine phase by 12.59 meV per formula unit (f.u.). In the $\sqrt{3}{\times}\sqrt{3}{\times}$2 CDW structure, the Sn1-Sn1 bond length $d_{\rm dimer}$ [see Fig. 1(b)] is shortened (lengthened) to be 3.052 or 3.073 {\AA} (3.710 or 3.740 {\AA}) compared to that (3.339 {\AA}) in the pristine structure. Hereafter, two Sn1 atoms between neighboring V$_3$Sn1 kagome layers are termed Sn dimers. While the CDW structure of AV$_3$Sb$_5$ has the large in-plane displacement of V kagome sublattice forming trimers and hexamers~\cite{AV3Sb5_chongze}, that of ScV$_6$Sn$_6$ has the minimal displacement of V atoms with the large out-of-plane displacement of Sc and Sn atoms~\cite{YBH_PRL_2023}. The variations of $d_{\rm dimer}$ during the CDW formation of ScV$_6$Sn$_6$ implies a charge transfer from the longer to the shorter Sn dimers, which in turn opens partial CDW gaps around the $H'$ = (${1 \over 3}$, 0, ${1 \over 4}$) point [see Fig. 2(b)]. Note that (i) the folded pristine band structure onto the $\sqrt{3}{\times}\sqrt{3}{\times}$2 supercell shows that the hybridized Sn1 $p_z$ and V $d_{xz,yz}$ states [$S_0$ in Fig. 2(c)] at $H'$ on the boundary of Brillouin zone have the six-fold degeneracy, which is split into the three occupied and three unoccupied states in the $\sqrt{3}{\times}\sqrt{3}{\times}$2 phase [see Fig. 2(b)] and (ii) the unfolded band structure of the $\sqrt{3}{\times}\sqrt{3}{\times}$2 phase shows little change along the ${\Gamma}$-$M$-$K$-${\Gamma}$ line around $E_F$ (see Fig. S5). Here, the six-fold degeneracy at $H'$ is generated by the folding of pristine bands along three different paths [see Fig. 2(c)]. As shown in Figs. 2(b) and 2(c), the charge character of the highest (lowest) occupied $S_1$ (unoccupied $S_2$) state at $H'$ represents the charge accumulation (depletion) in the shorter (longer) Sn1-Sn1 bonds, compared to the $S_0$ state in the pristine phase. Therefore, the $\sqrt{3}{\times}\sqrt{3}{\times}$2 CDW order can be attributed to a Jahn-Teller-like effect involving the Sn1-Sn1 bond distortions, the charge redistribution between such Sn dimers, and the resultant partial gap opening around the $H'$ point.

Table I shows the comparison of the lattice constants and $d_{\rm dimer}$ in the pristine phase of RV$_6$Sn$_6$ (R = Sc, Y, La). We find that the lattice constants $a$ = $b$ and $c$ in YV$_6$Sn$_6$ (LaV$_6$Sn$_6$) are larger than those in ScV$_6$Sn$_6$ by 0.84(1.91)\% and 0.14(0.63)\%, respectively. Meanwhile, YV$_6$Sn$_6$ (LaV$_6$Sn$_6$) has a much shorter $d_{\rm dimer}$ value of 3.151(3.051) {\AA}, compared to that (3.339 {\AA}) in ScV$_6$Sn$_6$. Note that the $d_{\rm dimer}$ values in the former compounds are close to the shortest one (3.052 {\AA}) in the $\sqrt{3}{\times}\sqrt{3}{\times}$2 CDW phase of ScV$_6$Sn$_6$. Therefore, we can say that the Jahn-Teller-like CDW formation with the Sn1-Sn1 bond distortions and their charge redistribution would not be necessary in YV$_6$Sn$_6$ and LaV$_6$Sn$_6$. According to the magnetization and x-ray diffraction measurements~\cite{jacs-Y-Lu} of Y- and Lu-doped ScV$_6$Sn$_6$ where Sc is substituted by larger Y or Lu, the CDW order was observed to be suppressed because the Sn1-Sn1 bond length is reduced by Y and Lu doping. It is thus expected that external perturbations such as chemical doping and pressure in RV$_6$Sn$_6$ can tune the Sn1-Sn1 bond length to suppress or create CDW orders.

\begin{table}[ht]
\caption{Calculated lattice constants and $d_{\rm dimer}$ in the pristine phase of RV$_6$Sn$_6$ (R = Sc, Y, La).}
\begin{ruledtabular}
\begin{tabular}{lccc}
   & $a$ = $b$ ({\AA}) & $c$ ({\AA}) & $d_{\rm dimer}$ ({\AA})   \\  \hline
ScV$_6$Sn$_6$  & 5.453 & 9.235 & 3.339  \\
YV$_6$Sn$_6$  & 5.499 & 9.248 & 3.151  \\
LaV$_6$Sn$_6$  & 5.557 & 9.293 & 3.051  \\
\end{tabular}
\end{ruledtabular}
\end{table}

Unlike the DFT prediction of the $\sqrt{3}{\times}\sqrt{3}{\times}$2 CDW ground state~\cite{YBH_PRL_2023}, experimental evidence~\cite{ScV6Sn6_CDW_PRL2022,exp1,exp2} showed that ScV$_6$Sn$_6$ undergoes a CDW transition to the $\sqrt{3}{\times}\sqrt{3}{\times}$3 phase at $T_{\rm CDW}{\approx}$92 K. Therefore, the temperature versus free energy landscape for the $\sqrt{3}{\times}\sqrt{3}{\times}$2 and $\sqrt{3}{\times}\sqrt{3}{\times}$3 phases can be illustrated like Fig. 3(a): i.e., the free energy of the $\sqrt{3}{\times}\sqrt{3}{\times}$3 phase becomes lower than that of the $\sqrt{3}{\times}\sqrt{3}{\times}$2 phase above $T_{\rm CDW}$. Our DFT calculations show that at zero temperature, the total energy of the $\sqrt{3}{\times}\sqrt{3}{\times}$3 phase is higher than that of the $\sqrt{3}{\times}\sqrt{3}{\times}$2 phase by ${\Delta}E$ = 6.29 meV/f.u. The $\sqrt{3}{\times}\sqrt{3}{\times}$3 CDW structure has two different kinds of the Sn1-Sn1 bonds with $d_{\rm dimer}$ = 3.169 and 3.677 {\AA} [see Fig. 1(b)]. Meanwhile, the band structure of the $\sqrt{3}{\times}\sqrt{3}{\times}$3 phase is overall the same as that of the $\sqrt{3}{\times}\sqrt{3}{\times}$2 phase (see Fig. S5), but the former has a CDW gap of 309 meV at the $H''$ = (${1 \over 3}$, 0, ${1 \over 3}$) point, slightly smaller than that (440 meV at $H'$) in the latter.

\begin{figure}[h!t]
\includegraphics[width=8.5cm]{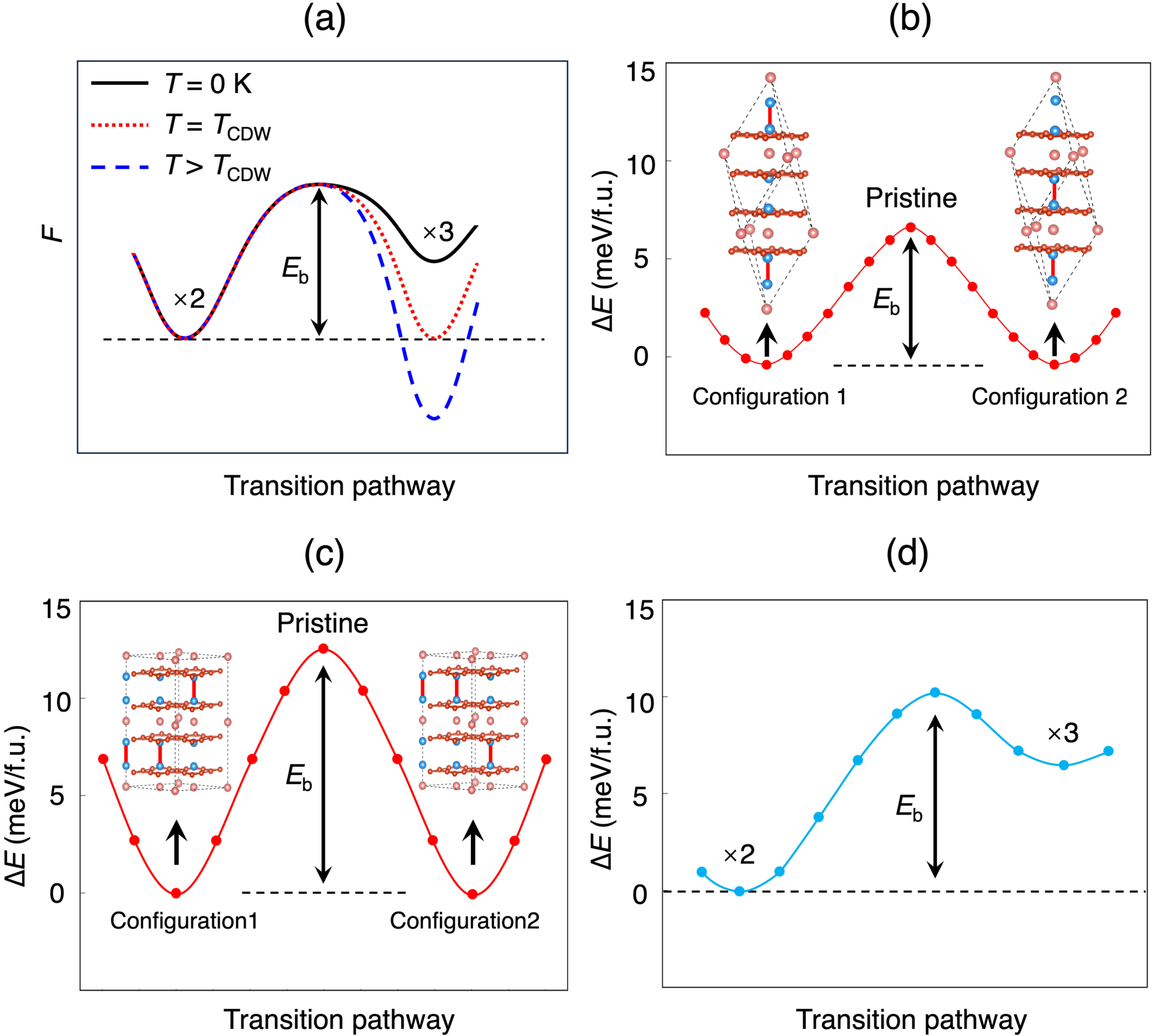}
\caption{(a) Schematic free energy landscape of the $\sqrt{3}{\times}\sqrt{3}{\times}$2 and $\sqrt{3}{\times}\sqrt{3}{\times}$3 phases at different temperatures and the energy profiles along the transition pathways (b) between degenerate $\sqrt{3}{\times}\sqrt{3}{\times}$3 configurations, (c) between degenerate $\sqrt{3}{\times}\sqrt{3}{\times}$2 configurations, and (d) between the $\sqrt{3}{\times}\sqrt{3}{\times}$2 and $\sqrt{3}{\times}\sqrt{3}{\times}$3 phases. The $E_b$ values in (b), (c), and (d) are estimated as 6.30, 12.59, and 10.24 meV/f.u., respectively.}
\label{figure:4}
\end{figure}

To estimate $T_{\rm CDW}$, we calculate the Helmholtz free energy difference ${\Delta}F$ = ${\Delta}E$ + ${\Delta}E_{\rm vib}$ - $T{\Delta}S$ between the $\sqrt{3}{\times}\sqrt{3}{\times}$3 and $\sqrt{3}{\times}\sqrt{3}{\times}$2 phases where $E_{\rm vib}$ represents vibrational energy~\cite{note-vib} and $S$ includes three different types of entropy such as vibrational entropy $S_{\rm vib}$, electronic entropy $S_{\rm el}$, and configurational entropy $S_{\rm con}$. $S_{\rm vib}$ is calculated from the phononic DOS $g^{\rm ph}(\epsilon)$~\cite{phononpy1,phononpy2}:
\begin{equation}
S_{\rm vib} = -3k_B \int g^{\rm ph}(\epsilon)(n(\epsilon){\rm ln}[n(\epsilon)] - [1+n(\epsilon)]{\rm ln}[1+n(\epsilon)]) d\epsilon,
\end{equation}
where $k_B$ is the Boltzmann constant and $n(\epsilon)$ is the Bose-Einstein population of a state of energy ${\epsilon}$ at $T$. Figure 4(a) displays ${\Delta}S_{\rm vib}$ = $S_{\rm vib,\sqrt{3}{\times}\sqrt{3}{\times}3}$ - $S_{\rm vib,\sqrt{3}{\times}\sqrt{3}{\times}2}$ as a function of $T$, indicating that the $\sqrt{3}{\times}\sqrt{3}{\times}$3 phase has larger $S_{\rm vib}$ than the $\sqrt{3}{\times}\sqrt{3}{\times}$2 phase. Therefore, $T{\Delta}S_{\rm vib}$ increases with increasing $T$ (see Fig. S6), leading to a decrease in ${\Delta}F$. Next, we calculate $S_{\rm el}$ from the electronic DOS $g^{\rm el}(\epsilon)$:
\begin{equation}
S_{\rm el} = -k_B \int g^{\rm el}(\epsilon)(n(\epsilon){\rm ln}[n(\epsilon)] + [1-n(\epsilon)]{\rm ln}[1-n(\epsilon)]) d\epsilon,
\end{equation}
where $n(\epsilon)$ is the Fermi-Dirac distribution function. Since the $\sqrt{3}{\times}\sqrt{3}{\times}$3 phase has higher $g^{\rm el}(E_F)$ than the $\sqrt{3}{\times}\sqrt{3}{\times}$2 phase (see Fig. S7), the former has larger $S_{\rm el}$ than the latter. However, due to the much smaller values of $S_{\rm el}$ [see Fig. 4(a)], $T{\Delta}S_{\rm el}$ plays a minor role in lowering ${\Delta}F$. Finally, we consider $S_{\rm con}$ arising from the number of degenerate configurations to which the system can be accessible. Using the nudged elastic band method~\cite{NEB1,NEB2}, we obtain an energy barrier $E_b$ = 6.30 meV/f.u. along the transition pathway TP$_1$ between degenerate configurations in the $\sqrt{3}{\times}\sqrt{3}{\times}$3 structure [see Fig. 3(b)], leading to strong CDW fluctuations above $T_{\rm CDW}$, as discussed below. Consequently, the $\sqrt{3}{\times}\sqrt{3}{\times}$3 phase can attain a configuration entropy of $S_{\rm con}$ = $k_B$ln(6)/3 = 0.051 meV$\cdot$K$^{-1}$/f.u. above $T_{\rm CDW}$ because there are six degenerate configurations within the primitive unit cell containing three formula units (see Fig. S8). On the other hand, the $\sqrt{3}{\times}\sqrt{3}{\times}$2 phase has relatively higher values of energy barrier [see Fig. 3(c)] and free energy [see Fig. 3(a)] above $T_{\rm CDW}$ compared to the $\sqrt{3}{\times}\sqrt{3}{\times}$3 phase, leading to $S_{\rm con}$ = 0 without dynamic fluctuations between degenerate configurations. Note that DS and IXS experiments~\cite{IXS_166,disorder1} observed the $\sqrt{3}{\times}\sqrt{3}{\times}$2 phase above $T_{\rm CDW}$ as a local short-range ordering. Given the estimated values of ${\Delta}E$, ${\Delta}E_{\rm vib}$, $T{\Delta}S_{\rm vib}$, $T{\Delta}S_{\rm el}$, $T{\Delta}S_{\rm con}$, we find that the $F$ of the high-entropy $\sqrt{3}{\times}\sqrt{3}{\times}$3 phase becomes equal to that of the $\sqrt{3}{\times}\sqrt{3}{\times}$2 phase at ${\sim}$97 K, where $T{\Delta}S_{\rm vib}$ = 1.03 meV/f.u., $T{\Delta}S_{\rm el}$ = 0.10 meV/f.u., $T{\Delta}S_{\rm con}$ = 4.99 meV/f.u., and ${\Delta}E_{\rm vib}$ = -0.17 meV/f.u.~\cite{note-vib} are obtained. This predicted $T_{\rm CDW}$ value is close to the experimental value of ${\sim}$92 K~\cite{ScV6Sn6_CDW_PRL2022} [see Fig. 4(b)], with highlighting the dominant role of configurational entropy in the $\sqrt{3}{\times}\sqrt{3}{\times}$3 CDW formation.

\begin{figure}[h!t]
\includegraphics[width=8.5cm]{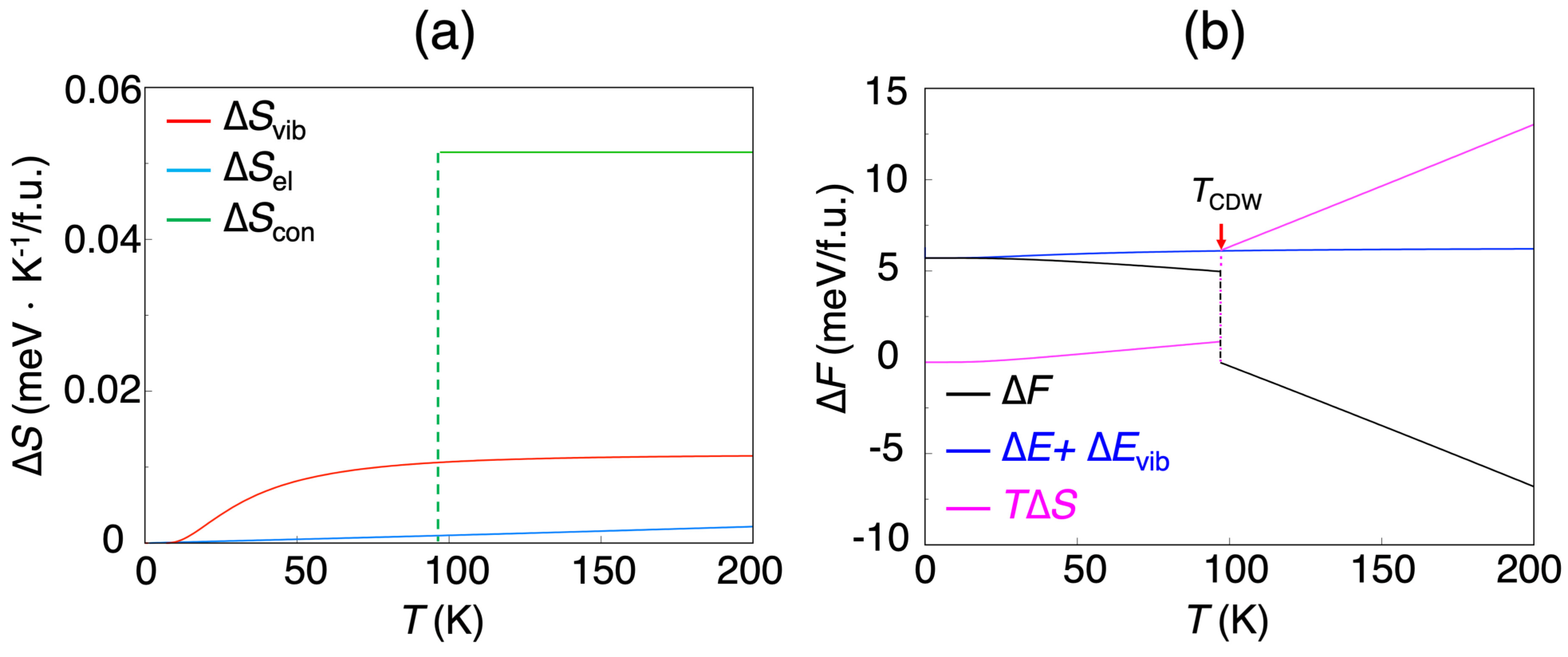}
\caption{Calculated ${\Delta}S$ and ${\Delta}F$. (a) ${\Delta}S_{\rm vib}$, ${\Delta}S_{\rm el}$, ${\Delta}S_{\rm con}$, and (b) ${\Delta}F$ as a function of $T$. In (b), the calculated values of ${\Delta}E$+${\Delta}E_{\rm vib}$ and $T{\Delta}S$ are also included.}
\label{figure:3}
\end{figure}

According to the Arrhenius equation~\cite{Ae}, the transition rate $R$ can be estimated as $R$ = $\nu$ exp($-{E_{b} \over k_{B}T}$), where $\nu$ is the attempt frequency and $E_b$ is the activation barrier. To estimate $R$ along TP$_1$, we adopt $\nu$ with the frequency (${\sim}$1.4 THz) of the Sn1-Sn1 bond stretching phonon modes around the $H$ point in the $\sqrt{3}{\times}\sqrt{3}{\times}$3 phase [see Fig. S4(c)]. With $E_b$ = 6.30 meV along TP$_1$, we obtain $R$ = 7.61$\times$10$^{11}$ s$^{-1}$ at 120 K above $T_{\rm CDW}$, indicating strong CDW fluctuations among the degenerate configurations of the $\sqrt{3}{\times}\sqrt{3}{\times}$3 phase with a characteristic lifetime
of 0.91 ps~\cite{half-life-time-note}. Interestingly, this fluctuating CDW lifetime is comparable with that (${\sim}$0.5 ps) measured by ultrafast spectroscopy
in cuprates~\cite{cuprate-half-life}. We note that such CDW fluctuations above $T_{\rm CDW}$ are similar to the case of CsV$_3$Sb$_5$ where the dynamic fluctuations of the 2${\times}$2${\times}$2 CDW order was experimentally observed above $T_{\rm CDW}$ = 94 K up to ${\sim}$160 K~\cite{fluctuations135}. Just below $T_{\rm CDW}$, the energy barrier from the $\sqrt{3}{\times}\sqrt{3}{\times}$3 to the $\sqrt{3}{\times}\sqrt{3}{\times}$2 phase is estimated as ${\sim}$10.24 meV [see Figs. 3(a) and 3(d)], relatively higher than that (6.30 meV) along TP$_1$. Therefore, the long-range $\sqrt{3}{\times}\sqrt{3}{\times}$3 phase can be kinetically trapped in certain temperature ranges below $T_{\rm CDW}$ without undergoing a transition to the $\sqrt{3}{\times}\sqrt{3}{\times}$2 phase, which is supported by experimental evidence that the $\sqrt{3}{\times}\sqrt{3}{\times}$3 phase was observed down to a temperature of ${\sim}$50 K~\cite{ScV6Sn6_CDW_PRL2022}. Furthermore, as shown in Fig. 4(b), a discontinuous change in ${\Delta}F$ with a sizable release of configurational entropy indicates the first-order character of CDW phase transition, consistent with the presence of a sharp peak in the measured specific heat data~\cite{ScV6Sn6_CDW_PRL2022} at $T_{\rm CDW}$. It is worth noting that, as $T$ further decreases, both the energy barrier from the $\sqrt{3}{\times}\sqrt{3}{\times}$3 to the $\sqrt{3}{\times}\sqrt{3}{\times}$2 phase and its thermal activation rate are simultaneously lowered [see Fig. 3(a)]. Consequently, these two factors represent the compensation effect that determines whether the $\sqrt{3}{\times}\sqrt{3}{\times}$3 phase remains kinetically trapped or is transformed into the $\sqrt{3}{\times}\sqrt{3}{\times}$2 phase. Indeed, synchrotron x-ray diffraction experiments~\cite{disorder1} observed the coexistence of the $\sqrt{3}{\times}\sqrt{3}{\times}$2 and $\sqrt{3}{\times}\sqrt{3}{\times}$3 charge correlations down to 15 K, indicating that the measured samples contain a mixture of short-range $\sqrt{3}{\times}\sqrt{3}{\times}$2 and long-range $\sqrt{3}{\times}\sqrt{3}{\times}$3 CDW orders at lower temperatures.

$Discussion$.$-$ Recently, IXS measurements~\cite{IXS_166} for ScV$_6$Sn$_6$ reported an unusual CDW formation process with two competing CDW instabilities: i.e., the long-range $\sqrt{3}{\times}\sqrt{3}{\times}$3 CDW below $T_{\rm CDW}$ while the short-range $\sqrt{3}{\times}\sqrt{3}{\times}$2 CDW above $T_{\rm CDW}$. Across the CDW phase transition, the growth of an elastic central peak corresponding to the $\sqrt{3}{\times}\sqrt{3}{\times}$3 CDW was observed together with phonon softening associated with the $\sqrt{3}{\times}\sqrt{3}{\times}$2 CDW. It is, however, a puzzle that the soft phonon occurring at $q^*$ = (${1 \over 3}$, ${1 \over 3}$, ${1 \over 2}$) is accompanied by the CDW formation at a different wavevector $q_s$ = (${1 \over 3}$, ${1 \over 3}$, ${1 \over 3}$). Furthermore, the maximum IXS peak intensity of the $\sqrt{3}{\times}\sqrt{3}{\times}$3 CDW is at least 3 orders of magnitude larger than that of the $\sqrt{3}{\times}\sqrt{3}{\times}$2 CDW~\cite{IXS_166}, implying that the former phase has significantly larger domain sizes in the measured samples than the latter one. Therefore, it is unlikely that there exists a first-order transformation from the short-range $\sqrt{3}{\times}\sqrt{3}{\times}$2 to the long-range $\sqrt{3}{\times}\sqrt{3}{\times}$3 CDW orders, driven by the scenario of a momentum-dependent electron-phonon coupling favoring $q_s$-CDW as the ground state~\cite{IXS_166}. Instead, we believe that the observed first-order transition is of an order-disorder type between the static $\sqrt{3}{\times}\sqrt{3}{\times}$3 CDW and its fluctuations, as discussed above. Here, the dynamic fluctuations of $\sqrt{3}{\times}\sqrt{3}{\times}$3 CDW above $T_{\rm CDW}$ should sharply weaken the intensity of elastic central peak at $q_s$, as measured by IXS~\cite{IXS_166}. Meanwhile, according to synchrotron x-ray diffraction measurements~\cite{disorder1}, the short-ranged charge correlations of the $\sqrt{3}{\times}\sqrt{3}{\times}$2 phase was observed in a wide range of temperatures between 15 and 300 K. This may be explained in terms of the local ordering character of the $\sqrt{3}{\times}\sqrt{3}{\times}$2 phase due to a relatively higher energy barrier between its degenerate configurations [see Fig. 3(c)]. More refined experimental identifications of the phase-dependent CDW fluctuations above $T_{\rm CDW}$ are needed in the future.

In conclusion, based on first-principles calculations, we have demonstrated that the CDW order in ScV$_6$Sn$_6$ is driven by a Jahn-Teller-like effect involving the Sn1-Sn1 bond distortions, charge redistribution, and partial band gap opening. Furthermore, our free energy analysis revealed that the configurational entropy in the $\sqrt{3}{\times}\sqrt{3}{\times}$3 phase emerges above $T_{\rm CDW}$ due to CDW fluctuations, while below $T_{\rm CDW}$, the long-range $\sqrt{3}{\times}\sqrt{3}{\times}$3 CDW order is kinetically trapped, indicating a first-order, order-disorder phase transition. Our findings clarify the driving mechanism of CDW and the order-disorder nature of CDW phase transition in ScV$_6$Sn$_6$, which have important implications for understanding the observed first-order transition with CDW fluctuations in other kagome metal compounds such as AV$_3$Sb$_5$~\cite{firstorder135-1,firstorder135-2,fluctuations135} and FeGe~\cite{FeGe}.

\vspace{0.4cm}

\noindent {\bf Acknowledgements.}
This work was supported by the National Research Foundation of Korea (NRF) grant funded by the Korean Government (Grants No. 2022R1A2C1005456 and No. RS202300218998) and by BrainLink program funded by the Ministry of Science and ICT through the National Research Foundation of Korea (2022H1D3A3A01077468). Z.Z. acknowledges Innovation
Program for Quantum Science and Technology (Grant No. 2021ZD0302800) and Y.J. acknowledges National Natural Science Foundation of China (Grant No. 12074099). The calculations were performed by the KISTI Supercomputing Center through the Strategic Support Program (Program No. KSC-2023-CRE-0112) for the supercomputing application research.  \\

S. L. and C. W. contributed equally to this work.

\vspace{0.4cm}

\noindent $^{*}$ Corresponding authors: zhangzy@ustc.edu.cn and chojh@hanyang.ac.kr


\begin{thebibliography}{99}
\bibitem{method} Our first-principles DFT calculations were performed using the Vienna $ab$ $initio$ simulation package with the projector-augmented wave method~\cite{vasp1,vasp2,paw}. For the exchange-correlation energy, we employed the generalized-gradient approximation functional of Perdew-Burke-Ernzerhof~\cite{pbe}. The plane wave basis was employed with a kinetic energy cutoff of 400 eV, and the $k$-space integration was done with 12${\times}$12${\times}$8, 6${\times}$6${\times}$4, and 8${\times}$8${\times}$8 meshes for the pristine, $\sqrt{3}{\times}\sqrt{3}{\times}$2, and $\sqrt{3}{\times}\sqrt{3}{\times}$3 phases, respectively. Here, the $\sqrt{3}{\times}\sqrt{3}{\times}$2 and $\sqrt{3}{\times}\sqrt{3}{\times}$3 structures were considered with the hexagonal and primitive rhombohedral unit cells, respectively [see Fig. 1(b)]. All atoms were allowed to relax along the calculated forces until all the residual force components were less than 0.001 eV/{\AA}. The unfolded electronic band structure was calculated using the Vaspkit software~\cite{vaspkit}. The phonon dispersions were calculated using the finite displacement method implemented in the phonopy software ~\cite{phononpy1,phononpy2}. To calculate an energy barrier along the transition pathway between the $\sqrt{3}{\times}\sqrt{3}{\times}$2 and $\sqrt{3}{\times}\sqrt{3}{\times}$3 phases, we used the nudged elastic band method~\cite{NEB1,NEB2} within the same $\sqrt{3}{\times}\sqrt{3}{\times}$6 supercell.
\bibitem{ref_note1} A combined study of angle-resolved photoemission spectroscopy (ARPES) and DFT of ScV$_6$Sn$_6$~\cite{ARPES166} showed that the DFT band structures of the saddle points originating from the V 3$d$ orbitals are consistent with the ARPES data. Recent optical reflection measurements of ScV$_6$Sn$_6$~\cite{opticalPRB} also reported that the interband transitions associated with the saddle points at the $M$ point are hardly affected across the CDW transition and that the observed and DFT-predicted optical conductivities are in good agreement with each other. Since the previous and present DFT band structures are nearly the same (see Fig. S2), we can say that our DFT calculations properly capture the single-particle excitation spectra measured by ARPES~\cite{ARPES166} and infrared spectroscopy~\cite{opticalPRB}
\bibitem{note-vib} The vibration energy is given by $E_{\rm vib} = \sum_{\mathbf{q} N} \hbar \omega_{\mathbf{q} N}\left[\frac{1}{2}+\frac{1}{\exp \left(\hbar \omega_{\mathbf{q} N} / k_{\mathrm{B}} T\right)-1}\right]$ where $\mathbf{q}$ and $N$ represent the wave vector and band index of phonon modes, respectively. The calculated result of ${\Delta}E_{\rm vib}$($T$) is displaced in Fig. S6.
\bibitem{half-life-time-note} The characteristic lifetime (half lifetime) of the transition between two degenerate $\sqrt{3}{\times}\sqrt{3}{\times}$3 configurations is calculated using the equation $t_{1/2}$ = ln(2)/$R$~\cite{kinetics}.
\end{thebibliography}
\end{document}